\documentclass[a4paper, 10pt]{article}

\usepackage[utf8]{inputenc}
\usepackage{authblk}
\usepackage{graphicx}
\usepackage{amsmath}
\usepackage{amssymb}
\usepackage{amsthm}
\usepackage{color}
\usepackage{url}
\usepackage{hyperref}
\usepackage{enumitem}
\usepackage{stmaryrd}
\usepackage{stmaryrd}
\SetSymbolFont{stmry}{bold}{U}{stmry}{m}{n} 
\usepackage{tikz}
\usepackage{setspace}
\usepackage{verbatim}
\usepackage{float}
\usepackage{physics}

\usepackage{caption}
\usepackage{subcaption}

\usepackage{algpseudocode}
\usepackage{algorithm}

\usepackage[
  a4paper,
  left   = 25mm,
  right  = 25mm,
  top    = 10mm,
  bottom = 10mm,
  includehead,
  includefoot
]{geometry}

\usepackage{tabularx}
\usepackage{booktabs}
\usepackage{makecell}

\title{Minimal Time Robust Control for Two Superconducting Qubits}

\date{}

\author[1]{Niril~George\thanks{ngeorge99@proton.me (corresponding author)}}
\author[2]{Joseph~L.~Allen}
\author[3,4,5]{Robert~Kosut}
\author[1]{Eran~Ginossar\thanks{e.ginossar@surrey.ac.uk}}

\affil[1]{School of Mathematics and Physics, University of Surrey, Guildford, United Kingdom}
\affil[2]{RBA Acoustics Ltd, 44 Borough Road, London, United Kingdom}
\affil[3]{SC Solutions, Inc., San Jose, CA, USA}
\affil[4]{Princeton University, Princeton, NJ, USA}
\affil[5]{Quantum Elements, Inc., Thousand Oaks, CA, USA}

\begin{document}
\maketitle

\begin{abstract}
High-fidelity quantum gates are crucial for achieving fault-tolerant quantum computing; however, decoherence significantly reduces gate fidelities during long operation times. Although optimal control techniques can theoretically minimize these operation times, they often neglect realistic uncertainties in system parameters. In this work, we demonstrate that by using robust optimal control strategies, the cross-resonance gate in superconducting systems can be operated within $64$~ns, achieving fidelities of $\mathcal{F} > 0.99$ while maintaining robustness against up to $10\%$ uncertainty in a single parameter. Alternatively, by extending the control time to $71$~ns, we achieve fidelities of $\mathcal{F} > 0.999$ with robustness against up to $3\%$ uncertainty. Our results identify the minimal control times attainable with experimentally feasible pulses and system parameters, as well as the maximum allowable static parameter error for high-fidelity operations. Furthermore, we demonstrate simultaneous robustness against both static and time-dependent errors by generating $100$~ns control pulses ($\mathcal{F} > 0.99$) that maintain robustness against $10\%$ static parameter error and time-dependent parameter fluctuations two orders of magnitude stronger than typical experimental noise. These findings demonstrate a viable open-loop strategy for implementing fast, high-fidelity quantum gates in the presence of realistic system uncertainties that would otherwise degrade conventional control pulses.
\end{abstract}

\newpage
\section{\label{Minimal_Introduction}Introduction}
Superconducting qubits are among the most promising candidates for quantum computing architectures \cite{blais_cavity_2004, blais_quantum-information_2007, gu_microwave_2017, kjaergaard_superconducting_2020}. A high-precision implementation of desired unitary operations is essential for effective quantum computation \cite{aharonov_fault-tolerant_2008}. Errors in quantum gates primarily stem from incoherent processes, characterized by the $T_{1}$ and $T_{2}$ time constants, and from unitary errors due to imperfect control protocols and implementations \cite{barends_superconducting_2014}. Reducing gate times can mitigate incoherent errors by making their effects negligible. However, a fundamental speed limit exists for gate operations \cite{goerz_quantum_2011, levitin_operation_2002, svozil_maximum_2005, carlini_time-optimal_2006, carlini_time-optimal_2007, ashhab_speed_2012, lee_dependence_2018, howard_implementing_2023}. While these speed limits are typically derived analytically for simple two-level systems, more complex multi-level systems require numerical methods to determine minimal control times, which are inherently longer than the theoretical speed limits~\cite{kirchhoff_optimized_2018}.

Optimal control techniques have been widely employed to improve the fidelity of desired quantum operations by optimizing system parameters and pulse shapes \cite{ginossar_protocol_2010, goerz_charting_2017, goetz_quantum_2016, cross_optimized_2015, theis_high-fidelity_2016, schondorf_optimizing_2018, liebermann_optimal_2016, theis_simultaneous_2016, machnes_tunable_2018}. These techniques can be extended to mitigate the effects of noise, control errors, and system uncertainties, forming what is known as robust optimal control (ROC). Sampling-based ROC, for example, involves the search for design variables that maximize either the average or the worst-case gate fidelity across a set of uncertain parameters \cite{kosut_robust_2013, dong_robust_2015,dong_sampling-based_2023}. More advanced techniques employ stochastic sampling where new uncertainty samples are generated at each iteration, thereby improving generalization and scalability to higher-dimensional errors \cite{wu_learning_2019, turinici_stochastic_2019}. Other strategies achieve robustness analytically, including composite pulse sequences that cancel amplitude and detuning errors \cite{wimperis_broadband_1994, cummins_tackling_2003}, dynamically corrected gates that embed the target operation in engineered pulse sequences to suppress decoherence and control imperfections \cite{khodjasteh_dynamically_2009, khodjasteh_automated_2012, buterakos_geometrical_2021, walelign_dynamically_2024}, and DRAG (derivative removal by adiabatic gate) and its variants, which use derivative-quadrature pulse shaping to mitigate leakage and phase errors \cite{motzoi_simple_2009,gambetta_analytic_2011,chen_measuring_2016, mckay_efficient_2017}. Machine learning has also been used to improve robustness, from simulation-based training \cite{shi_supervised_2024, sivak_model-free_2022, niu_universal_2019} to on-hardware training of control pulses \cite{baum_experimental_2021}. More advanced frameworks seek to optimize for both robustness and speed, shifting the focus to time-optimal robust control \cite{cao_robust_2024, zeng_fastest_2018, van_damme_robust_2017, hanson_constructing_2024, aroch_mitigating_2024}; however, extending these results to multi-qubit gates in superconducting circuits remains an open challenge.

The cross-resonance gate is one of the most effective superconducting quantum gates for generating entanglement \cite{rigetti_fully_2010, chow_simple_2011, magesan_effective_2020}. It is an all-microwave gate implemented on fixed-frequency qubits, leveraging the cross-coupling between them. This gate is advantageous for superconducting transmon qubits, which offer long coherence times \cite{reagor_quantum_2016}, reduced charge noise \cite{koch_charge-insensitive_2007}, and high single-qubit gate fidelities \cite{li_error_2023}, owing to the use of fixed-frequency transmons that can be engineered for optimal performance. The minimal overhead of the gate, requiring only microwave control, makes it particularly suitable for scaling up quantum computers.

Reported gate times for the cross-resonance gate have been relatively slow compared to those for flux-tunable gates. Published benchmarks include $\mathcal{F} = 0.9910 \pm 0.002$ at gate times of $160$~ns from Sheldon et al. \cite{sheldon_procedure_2016}, rising to $\mathcal{F} \approx 0.9917$ with error-mitigation techniques at $484$~ns \cite{sundaresan_reducing_2020}. System-wide characterizations of large-scale IBM processors report median two-qubit fidelities of $\mathcal{F} \in [0.976, 0.993]$ with gate times $t_g \in [460, 666]~\text{ns}$ \cite{abughanem_ibm_2025}, while in-situ deep reinforcement–learning optimization has improved the fidelities to $\mathcal{F} \approx 0.995$ with similar gate times on the same hardware \cite{baum_experimental_2021}. Cross-resonance–enabled two-qubit gates have also demonstrated $\mathcal{F} = 0.9977 \pm 0.0002$ in $180$~ns \cite{kandala_demonstration_2021}. While fidelities at this level are sufficient for fault-tolerant implementations within the surface-code architecture \cite{aharonov_fault-tolerant_2008, satzinger_realizing_2021, nielsen_quantum_2012}, there remains room for improvement in both fidelity and operation time. A time-optimal control approach was applied to the cross-resonance gate \cite{kirchhoff_optimized_2018}, revealing that significantly shorter gate times are possible: it numerically determined an unconstrained quantum speed limit of $5\text{--}10$~ns and identified experimentally feasible pulses achieving fidelity $\mathcal{F} \approx 0.999$ in $70$~ns with a bandwidth-constrained single-carrier drive. Robust optimal control has been applied to similar cQED gates, demonstrating multi-level model fidelities of $\mathcal{F} > 0.9639$ in $199$~ns robust to $\pm 1\%$ static parameter uncertainty \cite{allen_optimal_2017}. More recently, reinforcement learning approaches have generated pulses ($249$~ns) that can achieve $\mathcal{F} > 0.999$ with robustness to hardware drifts of up to $\pm 4\%$ in simulation \cite{nam_nguyen_reinforcement_2024}, while Baum et al. \cite{baum_experimental_2021} experimentally demonstrated that their deep reinforcement learning optimized gates maintained fidelities $\mathcal{F} > 0.993$ against calibration drifts for over 25 days.

In this work, we employ the robust optimal control methodology from~\cite{kosut_robust_2013} to ascertain the minimal gate time necessary for achieving high-fidelity operations while ensuring robustness against uncertainties in the system's intrinsic parameters. Our model encompasses the four lowest energy states of the transmon, effectively representing two directly coupled multi-level qubits with direct driving mechanisms. We investigate varying levels of static uncertainty in a single system parameter, spanning from $1\%$ to $10\%$, which are within the bounds of experimental relevance. Our findings reveal that control pulses with durations of $64$~ns maintain fidelities exceeding $0.99$ across all considered uncertainty levels. Additionally, pulses of $71$~ns achieve fidelities above $0.999$ for uncertainties up to $3\%$. Notably, securing a worst-case fidelity of $\mathcal{F} > 0.999$ within multi-level systems for gate times under $100$~ns constrains the maximum permissible uncertainty to $3\%$. This highlights an inherent trade-off between uncertainty and performance, underscoring the importance of fabricating hardware with well-characterized uncertainties, rather than merely augmenting control strategies in existing systems. We further show that our framework exhibits robustness to time-dependent errors by generating control pulses under $100$~ns achieving $\mathcal{F} > 0.99$ that maintain $10\%$ level of robustness against static parameter error while simultaneously establishing robustness against time-dependent parameter fluctuations. We confirmed this robustness not only in realistic noise simulations but also in strong noise simulations where parameter fluctuations are amplified by two orders of magnitude. Our results enable the application of robust optimal control techniques in future two-qubit gate experiments, demonstrating that pulse-shaping methods can effectively incorporate uncertainties to attain high fidelities with minimal gate durations.

The remainder of this paper is organized as follows: In Sec.~\ref{Minimal_CR}, we define the theoretical model for the cross-resonance gate used in this study. Sec.~\ref{Minimal_ROC} then describes the robust optimal control methods and the algorithm employed to achieve the reported results. Next, in Sec.~\ref{Minimal_NonRobust}, we determine the minimal time for the cross-resonance gate using experimentally feasible parameterizations without considering robustness. Following this, Sec.~\ref{Minimal_Robust} presents the results of the search for a time-optimal cross-resonance gate robust to static parameter uncertainty. In Sec.~\ref{Minimal_Noise}, we demonstrate that the robust cross-resonance gate operation remains effective under extreme conditions by establishing simultaneous robustness against static and time-dependent errors. Finally, Sec.~\ref{Minimal_Conclusion} presents our conclusions.

\section{\label{Minimal_CR}Cross-resonance gate for multi-level qubits}

In this work, we analyze the cross-resonance gate between two transmons coupled by a bus resonator, with microwave drives directly applied to each transmon, which replicates the experimental setup reported by Sheldon et al.~\cite{sheldon_procedure_2016}. When the resonator’s fundamental frequency is significantly higher than the $\ket{0} \leftrightarrow \ket{1}$ transition frequencies of the individual transmons, the Hamiltonian can be projected onto the zero-excitation subspace of the resonator. By additionally applying the rotating wave approximation, we obtained the following lab-frame effective Hamiltonian:
\begin{eqnarray}
\frac{H}{\hbar} = \sum_{j=1}^{2} \left( \omega_{j} b^{\dagger}_{j} b_{j} + \frac{\delta_{j}}{2} b^{\dagger}_{j} b_{j} (b^{\dagger}_{j} b_{j} - 1) \right)  + J (b^{\dagger}_{1} b_{2} + b_{1} b^{\dagger}_{2})
 + \sum_{j=1}^{2} \varepsilon_{j}^{k}(t) \cos(\omega_{j}^{k} t + \phi_{j}^{k}) (b^{\dagger}_{j} + b_{j}). 
 \label{Minimal_TwoCoupledTransmon}
\end{eqnarray}
Here, $b_{j}$ and $b^{\dagger}_{j}$ represent the annihilation and creation operators of transmon $j$, $\omega_{j}$ is the frequency of transmon $j$, $\delta_{j}$ is its anharmonicity, $J$ is the coupling strength between the two transmons, and $\varepsilon^{j}_{k}(t)$ is the pulse envelope (control) for drive $k$ on transmon $j$ with carrier frequency $\omega_{j}^{k}$ and phase $\phi_{j}^{k}$. We assumed that the Hamiltonian for the transmon is of the form of a Duffing oscillator \cite{cross_optimized_2015}, which is valid if the anharmonicity is in the transmon regime.

To perform a cross-resonance gate, the "control" qubit is driven by a direct microwave drive at the frequency of the "target" qubit \cite{chow_simple_2011,malekakhlagh_first-principles_2020}. Due to the cross-coupling between the qubits, this generates an entangling operation. To see this, consider a Hamiltonian of two directly coupled qubits with a single microwave drive on the first qubit, with control only in the x-quadrature:
\begin{eqnarray}
  \frac{H}{\hbar} = \frac{\omega_{1}}{2}\sigma_{z}^{(1)} + \frac{\omega_{2}}{2}\sigma_{z}^{(2)} + J\left(\sigma_{+}^{(1)}\sigma_{-}^{(2)} + \sigma_{-}^{(1)}\sigma_{+}^{(2)}\right) + \varepsilon(t)\sigma_{x}^{(1)}\cos(\omega_{d} t).
\end{eqnarray}
This Hamiltonian indicates that entanglement can be generated by the coupling between the qubits by simply waiting an appropriate amount of time. However, as $J$ is generally small, this time is very long. Instead, because the coupling between the qubits is weak, such that $J \ll |\Delta_{12}| = |\omega_{1} - \omega_{2}|$, we can diagonalize the first line of the Hamiltonian to effectively decouple the qubits. Using this transformation, the drive term can be transformed, so the full Hamiltonian becomes
\begin{eqnarray}
 \frac{ H_{q}'}{\hbar} = \frac{\omega_{1}'}{2}\sigma_{z}^{(1)}+\frac{\omega_{2}'}{2}\sigma_{z}^{(2)} +\varepsilon(t)\left(\sigma_{x}^{(1)}+\frac{J}{\Delta_{12}}\sigma_{z}^{(1)}\sigma_{x}^{(2)}\right)\cos(\omega_{d}t),
\end{eqnarray}\label{Minimal_CRHam}
with $\omega_{1}'=\omega_{1}+J^{2}/\Delta_{12}$ and $\omega_{2}'=\omega_{2}-J^{2}/\Delta_{12}$ being the decoupled, shifted qubit frequencies. Therefore, by choosing the drive frequency $\omega_{d}=\omega_{2}'$, a two-qubit entangling operation using $\sigma_{z}^{(1)}\sigma_{x}^{(2)}$ can be performed.

For simplicity, in Eq.~\ref{Minimal_CRHam}, only single-quadrature control was chosen. Generally, in experiments, there is full control of both quadratures for the drives. Additionally, for direct drives on the qubits, there are two microwave control lines, one for each qubit, as is the case in ref.~\cite{sheldon_procedure_2016} . In this case, in the drive frame, the Hamiltonian becomes
\begin{eqnarray} \label{Minimal_Equation_TwoQuadratureCoupledTransmons}
 \frac{H}{\hbar} &=& \sum_{j=1,2}\left(\Delta_{j}b^{\dagger}_{j}b_{j}+\frac{\delta_{j}}{2}b^{\dagger}_{j}b_{j}\left(b^{\dagger}_{j}b_{j}-1\right)\right) 
 +J\left(b^{\dagger}_{1}b_{2}+b_{1}b^{\dagger}_{2}\right) \nonumber \\
 && +\sum_{j=1,2}\left(\varepsilon^{x}_{j}(t)\left(b^{\dagger}_{j}+b_{j}\right)+i\varepsilon^{y}_{j}(t)\left(b^{\dagger}_{j}-b_{j}\right)\right), 
\end{eqnarray}
with $\Delta_{j}$ being the detuning of transmon $j$ from the drive frequency. For the simulations, we transformed the Hamiltonian into the dressed basis and truncated the matrix to include only the first four energy levels of both transmons.

Higher energy levels were excluded from our analysis, as our simulations (see Sec.~\ref{Minimal_Population} and Fig.~\ref{Minimal_Figure_Leakage}) showed that energy levels beyond the third had minimal occupation when the drive power for the cross-resonance drive was set to match the parameters in \cite{sheldon_procedure_2016}. Furthermore, we precisely matched the system parameters to those used in the referenced study for a fair comparison. Specifically, the frequencies were set at $\omega_{1}/2\pi = 4.914$\,GHz and $\omega_{2}/2\pi = 5.114$\,GHz, and both transmons exhibited an anharmonicity of $-330$\,MHz. The bus resonator frequency was maintained at $6.31$\,GHz, resulting in an estimated coupling of $J/2\pi = 3.8$\,MHz between the two qubits.

\section{\label{Minimal_ROC}Robust Optimal Control}

Many experimental procedures have refined the use of standard pulse shapes by calibrating the pulses to minimize errors due to drifts in the control fields or the inevitable filtering they undergo. However, as these approaches reach the limits of what is achievable with standard pulses and current hardware, there is a limit on the achievable gate time, as a standard pulse requires a certain duration to perform the desired gate. Quantum optimal control provides a method for creating shaped, intelligent pulses that can allow for shorter gate times. However, these are often limited in their effectiveness, as the optimal control simulation requires a predefined model with specific system parameters.

It is acknowledged that parameter estimation in multi-qubit superconducting systems requires improvement \cite{stenberg_efficient_2014, bejanin_resonant_2020}. Previously, it has been shown that if parameter uncertainty is not included in the optimization, the fidelity over the uncertain range significantly decreases away from the simulated parameter value \cite{kosut_robust_2013, allen_optimal_2017}. Measuring system parameters is resource-intensive, and as the scale of quantum computers increases, this becomes a significant issue. As systems scale up, this leads to greater uncertainty in the system parameters and therefore likely gives rise to optimal control solutions that, when applied experimentally, achieve significantly lower fidelity than simulated. Robust optimal control seeks to ensure that this drop-off in fidelity is reduced considerably by including robustness in the search. This allows optimal-control-shaped pulses to be less limited by the constraints of the simulated model and to provide experimentally feasible results. Robustness aids in scaling up quantum computers by ensuring less variability in the performance of quantum gates across a multi-qubit processor and making the implementation more reliable overall. The work in this paper used numerical optimization techniques to find optimal pulse shapes to perform the gate of interest in minimal time. In particular, the interest was in achieving fast gate times by finding pulses that are robust to uncertainties. By decreasing the times, incoherent errors, such as decoherence, can be effectively removed. In this case, the focus was on the coherent errors resulting from, for example, parameter uncertainty.

Sampling-based robust optimization takes a sample of points, $\delta_{i}$, from some range, $\Delta$, of the uncertain parameters (or errors) \cite{kosut_robust_2013}. The fidelity is calculated from the time evolution of the Hamiltonian for each of the sampled parameter values, resulting in a range of fidelities, $\mathcal{F}_{i}$. Using this range of fidelities, the aim is then to either maximize the average fidelity of the range or to maximize the worst-case fidelity in the uncertain range \cite{kosut_robust_2013}. In general, for quantum computing, maximizing the average fidelity over the uncertain range is not enough. If the average fidelity is maximized, there may exist a value in the range of uncertainty that performs far below the desired threshold. In this case, if the real system parameter has a value at this sub-optimal point, then the actual performance of the gate would also be sub-optimal. Indeed, in this situation, slow parameter drifts may often encounter adverse parameter values and thus degrade the fidelity over the duration of a quantum algorithm. If, instead, the worst-case fidelity is maximized for the uncertain range, it is guaranteed that all values in the range perform at least as well as this worst-case fidelity, ensuring that slow parameter drifts do not adversely affect the fidelity during operation. Therefore, maximizing the worst-case fidelity is the ideal target for quantum gates.

The numerical optimization algorithm used for this work was based on the Sequential Convex Programming (SCP) algorithm \cite{kosut_robust_2013}. This is a gradient-based, local search optimizer, similar in concept to GRAPE \cite{khaneja_optimal_2005}. The SCP algorithm produces piecewise-constant amplitudes over $N$ uniform time intervals, $\tau = T/N$, where $T$ is the final time of the system evolution. The Hamiltonian at time $t_{k} = k\tau$ is then given by:
\begin{eqnarray}
    H(t_{k}) = H_{0} + \sum_{j} \varepsilon_{j}(t_{k}) H_{j}, \quad t_{k} = k \tau,
\end{eqnarray}
where $H_{0}$ is the drift (i.e., not controlled)) Hamiltonian, $j$ is the index of the control, $\varepsilon_{j}(t_{k})$ is the constant amplitude at time $t_{k}$, and $H_{j}$ is the time-independent control Hamiltonian associated with control $j$. Using the piecewise-constant approximation, the unitary evolution of the system $U(T)$ is given by $U(T) = \prod_{k=1}^{N} U_{k}$, with $U_{k} = \mathrm{exp}\big(-i H(t_{k}) \tau\big)$. In the piecewise-constant approximation, as the number of time intervals grows, so does the dimension of the control space. The optimization goal is to maximize a cost function of the form:
\begin{eqnarray}
    \mathcal{F} = \left| \frac{1}{n_{s}} \mathrm{Tr} \left(W^{\dagger} \hat{O} U(T) \hat{O} \right) \right|^{2},
\end{eqnarray}
with $\hat{O}$ being the projector into the computational subspace, $n_{s}$ the size of the computational subspace, $U(T)$ the time evolution of the full system at time $T$, and $W$ the desired unitary operator. This desired unitary operator was chosen as it is related to the CNOT operation via:
\begin{eqnarray}
    W = e^{-i \frac{\pi}{4} \sigma_{z}^{(1)} \sigma_{x}^{(2)}}, \quad \mathrm{CNOT} = e^{i \frac{\pi}{4} \sigma_{z}^{(1)}} 
    W 
    e^{i \frac{\pi}{4} \sigma_{x}^{(2)}}.
\end{eqnarray}
The optimizer is initialized by choosing an initial control, $\varepsilon$, which is the set of all complex-valued amplitudes at each time step. This initial control is within the feasible set $E$, defined by user-defined constraints on the control. The trust region, $\tilde{E}_{trust}$, effectively tells the optimizer the largest change that can be made to each individual time-step pulse while respecting the local linearization for the convex optimization update. Specifically, the trust region is adjusted after each optimization step: if the actual fidelity increases, the region becomes larger, whereas if the actual fidelity does not increase, the region is reduced to increase the likelihood of finding a change that enhances the fidelity. The uncertain set, $\Theta$, is created to define the size of the static uncertainty in the parameters of interest. From this uncertain set, a fixed sample of uncertain values, $\{\theta_i\} \subset \Theta$, is taken. At each iteration of the optimizer, SCP seeks to find an optimal solution for the increment, $\tilde{\varepsilon}$, by maximizing the linearized fidelities, $\mathcal{F}(\varepsilon,\theta_{i}) + \tilde{\varepsilon} \nabla_{\varepsilon}\mathcal{F}(\varepsilon,\theta_{i})$, subject to $\varepsilon + \tilde{\varepsilon} \in E$ and $\tilde{\varepsilon} \in \tilde{E}_{trust}$. Upon the return of the optimal $\tilde{\varepsilon}$ that satisfies all the constraints, the new fidelities are calculated and compared with the fidelities from the previous iteration. If $\mathrm{min}_{i}\mathcal{F}(\varepsilon+\tilde{\varepsilon},\theta_{i}) > \mathrm{min}_{i}\mathcal{F}(\varepsilon,\theta_{i})$, the control is updated as $\varepsilon \rightarrow \varepsilon + \tilde{\varepsilon}$, the trust region is increased, and the optimization step begins again. If, however, $\mathrm{min}_{i}\mathcal{F}(\varepsilon+\tilde{\varepsilon},\theta_{i}) < \mathrm{min}_{i}\mathcal{F}(\varepsilon,\theta_{i})$, the trust region is instead decreased, and the optimization step is repeated with the original $\varepsilon$. 

The key to the SCP algorithm is that the step to find the increment, $\tilde{\varepsilon}$, is a convex optimization. In addition, the algorithm compares the worst-case fidelity across the uncertain range at each iteration to maximize it. In the GRAPE algorithm, the method for increasing the fidelity is to update the control as $\varepsilon_{k} \rightarrow \varepsilon_{k} + \beta \frac{\partial \mathcal{F}}{\partial \varepsilon_{k}}$, where $k$ denotes the relevant piecewise-constant amplitude in the control and $\beta$ is a predefined increment. As the control is directly changed by the inclusion of the gradient, the only way to include robustness is to calculate the average fidelity across the uncertain range, e.g., $\mathcal{F}(\varepsilon) = \sum_{i=1}^{L} \mathcal{F}(\varepsilon, \theta_{i})$, and then use this for the gradient calculation and the update step. The update step then guarantees that the average fidelity over the uncertain range is maximized, not the worst-case fidelity. Other works have discussed sampling-based optimization for robust control by maximizing the average fidelity \cite{cykiert_robust_2024}. 

Alongside static parameter errors, superconducting qubits are also subject to time-dependent fluctuations in both the drift and control Hamiltonians. These arise from low-frequency drifts and $1/f$ noise in the device parameters~\cite{schreier_suppressing_2008,van_harlingen_decoherence_2004,schlor_correlating_2019,paladino_1_2014,tripathi_modeling_2024}, as well as broadband noise and imperfections in the classical control electronics~\cite{zurich_instruments_ag_hdawg_2022,gavrielov_spectrum_2025,gely_situ_2024,kumar_phase_2024,rohde_oscillator_2000,yakimov_nature_2020,heng_estimating_2024}, all of which can significantly impact gate fidelity. Consequently, to ensure high-fidelity operation, the robust optimal control framework must exhibit simultaneous robustness against both static and time-dependent errors.

This robustness is achieved by incorporating an ensemble of time-dependent errors trajectories, sampled from the underlying error statistics, directly into the cost function:
\begin{equation}
\overline{\mathcal{F}}(\varepsilon,\theta_i)
= \frac{1}{L}\sum_{l=1}^{L}
\mathcal{F}\!\big(\varepsilon,\,\theta_i,\,\boldsymbol{\beta}^{(l)}(t)\big).
\label{Minimal_NoiseMinMax}
\end{equation}
Here $L$ independent trajectories $\{\boldsymbol{\beta}^{(l)}(t)\}_{l=1}^{L}$ are generated for each static sample $\theta_i$, and they modulate the relevant Hamiltonian parameters in time. The robust objective is then taken as $\min_{i}\,\overline{\mathcal{F}}(\varepsilon,\theta_i)$, so that the optimisation targets the worst-case fidelity over the static error set, with each static point evaluated in terms of its average performance under time-dependent errors. The finite set of static samples $\{\theta_i\}$ is selected once at the beginning of the optimisation and kept fixed throughout, whereas the time-dependent errors trajectories $\boldsymbol{\beta}^{(l)}(t)$ are regenerated at every SCP iteration. This resampling prevents overfitting to any particular error realisation and drives the search towards pulses whose performance reflects the underlying error statistics. By choosing $L$ to be sufficiently large, the variance of this stochastic estimate is kept small enough for stable trust-region updates, yielding control pulses that are simultaneously robust to static parameter errors and to realistic time-dependent errors.

\section{\label{Minimal_NonRobust}Non-Robust Time Optimization}

\begin{table}[t!]
\caption{\textbf{Optimization Hyperparameters.} This table details the core parameters for the Sequential Convex Programming (SCP) algorithm, the stopping conditions for convergence, and the specific time grid and sampling strategies used for the non-robust (Sec.~\ref{Minimal_NonRobust}), static robust (Sec.~\ref{Minimal_Robust}), and simultaneous robust (Sec.~\ref{Minimal_Noise}) optimization scenarios. The term \textit{Dynamic} indicates that the stochastic error realizations are refreshed at each iteration.}
\label{tab:scp_hyperparameters}
\begin{tabularx}{\textwidth}{l X} 
\toprule
\multicolumn{2}{l}{\textbf{Core Algorithm Parameters}} \\
\midrule
Algorithm & Sequential Convex Programming \\
Random Initializations & 25 \\
Initial Trust Region  & 0.001 \\
Trust Region Increase Factor & 1.15 \\
Trust Region Decrease Factor & 0.5 \\
\midrule
\multicolumn{2}{l}{\textbf{Stopping Conditions}} \\
Max Iterations & 30000 \\
Fidelity Convergence  & $1 \times 10^{-9}$ over 10 iterations\\
Trust Region Convergence  & $1 \times 10^{-9}$ over 10 iterations \\
\midrule
\multicolumn{2}{l}{\textbf{Time Grid}} \\
Non-Robust Optimization & $1$–$100$ (1 ns steps) \\
Static Robust Optimization  & $1$–$100$ (1 ns steps) \\
Simultaneous Robust Optimization & 100 ns \\
\midrule
\multicolumn{2}{l}{\textbf{Sampling Parameters}} \\
Non-Robust Optimization & 1 Error Free Hamiltonian \\
Static Robust Optimization  & 3 Static Hamiltonians ($-\Delta J_{max}$, 0, $+\Delta J_{max}$) \\
Simultaneous Robust Optimization & 3 $\times$ (100 Dynamic Noisy + 1 Static Noise-Free) Hamiltonians \\
\bottomrule
\end{tabularx}
\end{table}

Before investigating time optimization for robust optimal control, it is useful to examine time minimization without robustness. As in almost all quantum control problems, solutions to time-optimization formulations are local. There are a variety of methods for finding local minimal time solutions. In simpler systems, a quantum speed limit can be derived using analytical methods. These speed limits are unique to the system of interest and the desired operation. However, for the multi-level systems in question, numerical methods are required to find the minimal operation time, for example, by including the gate time in the cost function to penalize long gate times ~\cite{dalessandro_introduction_2021}, or by plotting fidelity as a function of time ~\cite{kirchhoff_optimized_2018}. We adopted the latter strategy, running multiple optimizations from random initializations over a range of pulse durations to identify the shortest times required for high-fidelity operations.

Implementing the fidelity mapping required careful consideration of practical experimental constraints. While state-of-the-art AWGs have sampling rates of $50\,\text{GS/s}$, typical experimental setups are operated with AWGs that have sampling rates of $1\text{--}10\,\text{GS/s}$. Accordingly, we formulate the optimization over latent variables $c(t)$ and obtain the physical drive by Gaussian spectral shaping, described mathematically as $ \varepsilon(t)=\mathcal{F}^{-1}\!\big[G(\omega)\,\mathcal{F}[c(t)]\big]$.
Here, $\varepsilon(t)$ is the final pulse envelope applied to the hardware, $c(t)$ denotes the underlying control waveform optimized by the algorithm, $\mathcal{F}$ and $\mathcal{F}^{-1}$ are the Fourier transform and its inverse, respectively, $G(\omega)$ is a Gaussian low-pass filter in angular frequency $\omega$ (enforcing a prescribed bandwidth), and the product $G(\omega)\,\mathcal{F}[c(t)]$ represents the filtered Fourier spectrum of the optimized control. This procedure acts as a built-in low-pass that constrains the pulse bandwidth, mitigates high-frequency leakage to unwanted transitions, and ensures sufficient smoothness for accurate reproduction by the hardware. In our work, the filtering was set to have a minimum effective time step of $0.25\,\text{ns}$ (larger for longer pulses), which is compatible with standard AWG capabilities.

\begin{figure*}[!t]
\centering
\includegraphics[width=0.99\textwidth]{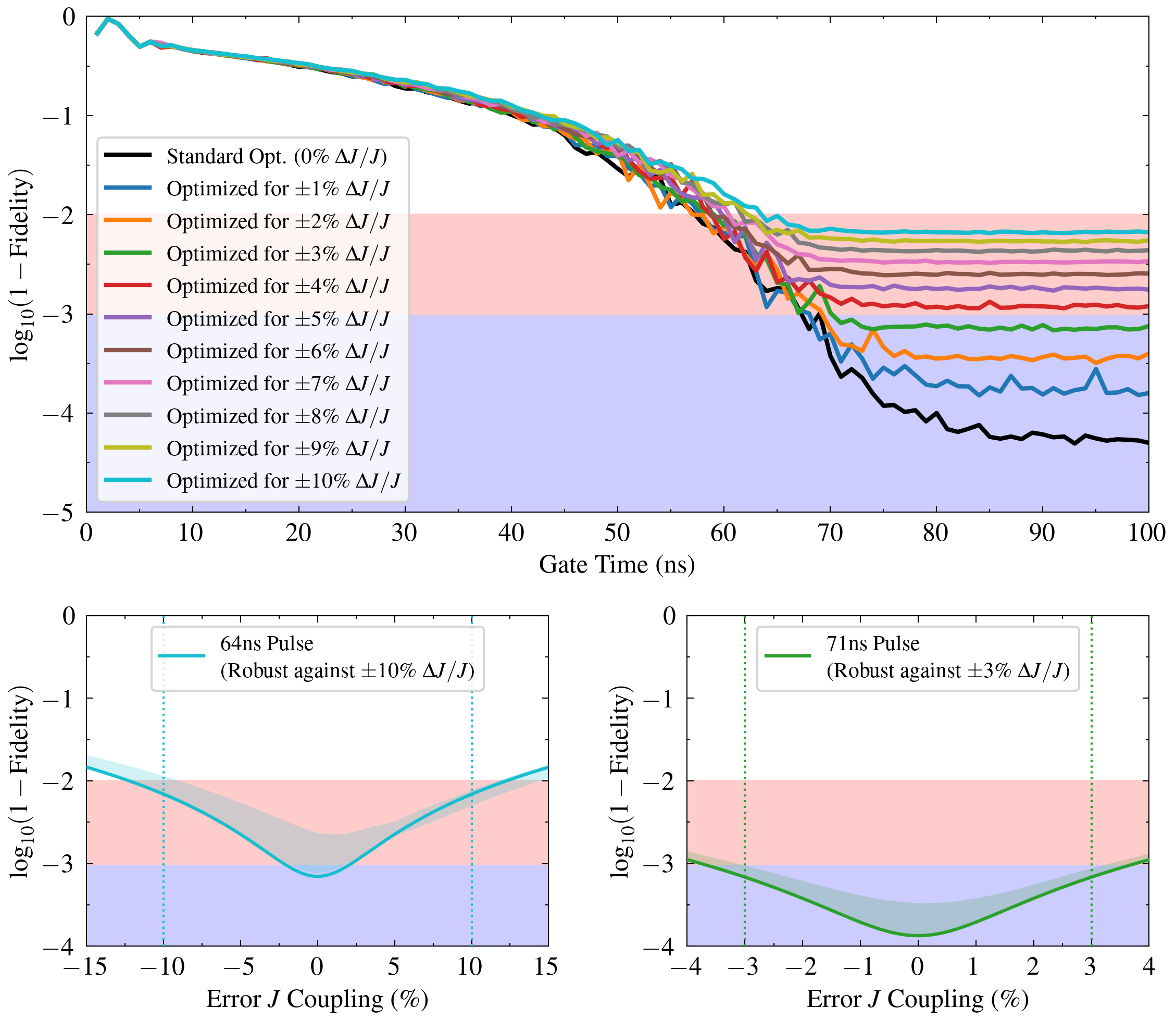}
\caption{
\textbf{Robust Cross-Resonance Gates Performance.}
    The top panel plots the cross-resonance gate infidelity ($\log_{10}(1 - \mathcal{F})$) achieved by the SCP optimization algorithm as a function of the target gate time ($1\,\text{ns}$ to $100\,\text{ns}$), where each curve corresponds to an optimization targeting robustness against a specific maximum static uncertainty in the $J$ coupling ($\Delta J/J$), ranging from 0\% (standard optimization, black line) up to $\pm 10\%$ (colored lines, see legend). For each gate time and uncertainty target, the optimization was run multiple times with different random initial pulses; the plotted infidelity at each $T$ is the average over random starts of the worst-case fidelity across the three sampled couplings. Background shading indicates target fidelity regions: $\mathcal{F} > 0.99$ (red, infidelity $< 10^{-2}$) and $\mathcal{F} > 0.999$ (light blue, infidelity $< 10^{-3}$). The minimum pulse duration for $\mathcal{F} > 0.99$ fidelity (optimized for $\pm 10\%$ robustness) is $64\,\text{ns}$. For $\mathcal{F} > 0.999$ fidelity (optimized for $\pm 3\%$ robustness), the minimum duration is $71\,\text{ns}$.
    The bottom panels show the infidelity ($\log_{10}(1 - \mathcal{F})$) of two representative pre-optimized pulses when subjected to a range of errors in the $J$ coupling: the left panel shows the performance of a pulse optimized for a $T = 64\,\text{ns}$ gate time targeting $\pm 10\%$ robustness in $\Delta J/J$, achieving the desired $\mathcal{F} > 0.99$ fidelity within this uncertainty range, while the right panel shows a pulse optimized for $T = 71\,\text{ns}$ targeting $\pm 3\%$ uncertainty, achieving $\mathcal{F} > 0.999$ within its range. The solid line indicates the infidelity of the best optimized pulse simulated at the corresponding coupling error, while the shaded area represents the infidelity range across an ensemble of optimized pulses (from the other starting points) tested at that error. The vertical dotted lines mark the $\pm$ error range for which each pulse was specifically optimized.}
\label{Minimal_Figure_Robust}
\end{figure*}

With these parameters defined, optimization was performed for pulse durations ranging from $1\,\text{ns}$ to $100\,\text{ns}$ in $1\,\text{ns}$ increments. For each duration, the SCP algorithm was executed with 25 independent random initial guesses, terminating upon reaching the maximum iteration limit or achieving convergence (see Table~\ref{tab:scp_hyperparameters} for all optimization parameters). The results of these standard (non-robust) optimizations are plotted in Figure~\ref{Minimal_Figure_Robust}, where the black line shows the average gate infidelity ($\log_{10}(1 - \mathcal{F})$) as a function of pulse duration. These optimizations yield average fidelities of $\mathcal{F} = 0.99$ in just $58\,\text{ns}$, $\mathcal{F} = 0.999$ in $68\,\text{ns}$, and $\mathcal{F} = 0.9999$ in only $79\,\text{ns}$, representing a twofold reduction in gate time for similar fidelity levels compared to current experimental demonstrations \cite{baum_experimental_2021,sheldon_procedure_2016, sundaresan_reducing_2020, abughanem_ibm_2025, kandala_demonstration_2021} and is notably consistent with the $70\,\text{ns}$ time reported in previous time-optimal (non-robust) studies of similar systems \cite{kirchhoff_optimized_2018}.

\section{\label{Minimal_Robust}Robust Time Optimization}

To provide an example of our robust optimal control framework and its effect on the minimal control time, a simplified model was considered where only a single static uncertainty was included in the system parameters. Typically, most system parameters in a two-qubit coupled system are easily measured using standard qubit spectroscopy with relatively low uncertainty. However, the qubit-qubit coupling cannot be directly measured using such methods. Instead, it is deduced via fitting to measurements~\cite{ku_suppression_2020}. For example, one method involves quantum state preparation of a control qubit (which introduces some error), followed by spectroscopic measurements on the target qubit, and then modeling of the coupling based on electromagnetic simulations of the modes. This leads to the qubit-qubit coupling $J$ being the most uncertain system parameter, and it was therefore the focus of this work.

To incorporate robustness into the time optimization, we adapted the optimization procedure outlined in Sec.~\ref{Minimal_NonRobust} with the modified objective to maximize the worst-case gate fidelity across a specified uncertainty range in the selected Hamiltonian parameters. As before, the SCP algorithm was executed for pulse durations from $1\,\text{ns}$ to $100\,\text{ns}$ (utilizing 25 independent initializations for each duration, as detailed in Table~\ref{tab:scp_hyperparameters}), for varying sizes of uncertainty in the coupling strength $J$, ranging from no uncertainty up to $\pm10\%$ error, in increments of $1\%$. For each uncertainty range, we sampled three values (which were kept fixed throughout the optimization): the maximum positive error, zero error, and the maximum negative error. Keeping the zero-error Hamiltonian is essential to ensure that we do not end up with a solution that has reduced fidelity at the most probable point. It is worth mentioning that when the error range is small, sampling the zero-error point may not be necessary~\cite{le_scalable_2023}. However, in a larger error range, it may be beneficial to sample more points~\cite{dong_robust_2015}.

The results of the robust optimization in Figure~\ref{Minimal_Figure_Robust} show that pulses optimized for robustness against a $\pm10\%$ uncertainty in the coupling strength $J$ can achieve a worst-case fidelity of $\mathcal{F} > 0.99$ in just $64\,\text{ns}$. Furthermore, by constraining the uncertainty to $\pm3\%$, worst-case fidelities above $\mathcal{F} = 0.999$ are attainable in $71\,\text{ns}$ (an example of this time-optimal robust pulse is shown in Fig.~\ref{Minimal_Figure_Robust_Pulse}).
Although this is $3\,\text{ns}$ slower than the non-robust case, it remains twice as fast as current experimental demonstrations \cite{baum_experimental_2021, sheldon_procedure_2016, sundaresan_reducing_2020, abughanem_ibm_2025, kandala_demonstration_2021}. These results also represent an improvement over previously reported robust control strategies for similar systems. For instance, we achieved a worst-case $\mathcal{F} > 0.999$ with $\pm 3\%$ robustness $3\times$ faster than reinforcement learning approaches that achieved average fidelities ($\mathcal{F} > 0.999$) against $\pm 4\%$ static drift in $249\,\text{ns}$ \cite{nam_nguyen_reinforcement_2024}. While at $\mathcal{F} > 0.99$, we achieved a much larger uncertainty robustness of $\pm 10\%$.

The bottom panels of Figure~\ref{Minimal_Figure_Robust} validate the effectiveness of the generated pulses by testing two example pulses against variations in the $J$ coupling. The left panel confirms that the $64\,\text{ns}$ pulse (optimized for $\mathcal{F} > 0.99$ with $\pm10\%$ robustness) successfully maintains its target fidelity across the entire error range. Similarly, the right panel shows that the $71\,\text{ns}$ pulse (designed for $\mathcal{F} > 0.999$ with $\pm3\%$ robustness) achieves its fidelity goal within the specified $\pm3\%$ window. At these short gate times, decoherence effects are negligible, as transmon coherence times are orders of magnitude longer---typically tens of microseconds~\cite{sheldon_procedure_2016} and up to $0.5\,\text{ms}$ for state-of-the-art devices~\cite{bland_2d_2025}.

We observed that there is a trade-off between gate operation time and static uncertainty. This can be understood as arising from an equivalence between the time constraint and the robustness requirement, as illustrated by the following simplified argument. Given a general Hamiltonian $H_0$ with a control $c(t)H_1$, the equation of motion is:
\begin{eqnarray}
\frac{dU(t)}{dt} = -i \left(H_0 + c(t) H_1\right) U(t).
\end{eqnarray}
When we reduce the time duration of the control task by rescaling time $t = (1-p)t^\prime$ with $0 < p \ll 1$, this is equivalent to a perturbation of the Hamiltonian, because:
\begin{eqnarray}
\frac{dU(t')}{dt'} = -i \left(H_0' + c(t') H_1'\right) U(t'),
\end{eqnarray}
where $H_0' = (1-p)H_0$ and $H_1' = (1-p)H_1$. In a situation where fidelity is locally a convex function of the two constraints, the fidelity may be maintained by compensating for the change in the time constraint with a change in uncertainty. Related ideas were discussed regarding the trade-off between fidelity and time-optimal control~\cite{moore_tibbetts_exploring_2012}. The results also indicate that the fidelity values do not increase linearly with gate operation time and that there appears to be a maximum reachable fidelity value, whereby an increase in gate operation time will not yield higher fidelity. For uncertainties greater than $3\%$, the best achievable worst-case fidelities are no greater than $\mathcal{F} = 0.999$, within the constraints of the simulated system. This suggests that the optimal level of uncertainty for this system is $3\%$ or less; if this level of measurement uncertainty for the $J$ coupling can be reached, then high-fidelity, short, robust pulses can be achieved.

It should be noted that to achieve robust and high-fidelity results, the transfer function of the AWGs, transmission lines, connectors, and other electronics should be included in the optimization conducted for a specific experiment. This will counteract a possible reduction in fidelity caused by transfer function distortions~\cite{allen_optimal_2017}.

\section{\label{Minimal_Noise}Robust Control against Time Dependent Errors}
To demonstrate simultaneous robustness against static errors and time-dependent errors, we subject the Hamiltonian parameters of the cross-resonance gate to stochastic variations. These variations simulate the effect of environmental noise and control imperfections under both realistic conditions and strong noise regimes, where fluctuations are amplified by two orders of magnitude. To systematically implement these fluctuations, we model each fluctuations, $\beta(t)$, as a stochastic process constructed from a sum of $N$ Fourier components \cite{shinozuka_simulation_1971}:
\begin{equation}
\beta(t) = \sum_{k=1}^{N} a_{k} \cos(\omega_{k} t + \phi_{k}) \label{eq:beta_definition_appendix}
\end{equation}
The spectral character of the perturbation is determined by the distributions for the amplitudes $a_k$ and frequencies $\omega_k$, allowing us to simulate common noise types. For each noise realization, the phases $\phi_{k}$ are drawn uniformly from $[0, 2\pi]$, and the amplitudes $a_k$ are initially drawn from a zero-mean Gaussian distribution. To generate $1/f^\alpha$ (flicker) noise, these amplitudes are then scaled by $\omega_k^{-\alpha/2}$, with frequencies $\omega_k$ sampled from a logarithmic distribution over the frequency band $[\omega_{\text{lower}}^{(1/f)}, \omega_{\text{upper}}^{(1/f)}]$. For white noise, the amplitudes are used as-is (unscaled), and the frequencies $\omega_k$ are sampled from a uniform distribution over $[\omega_{\text{lower}}^{\text{(w)}}, \omega_{\text{upper}}^{\text{(w)}}]$. This process models the physical effects of fluctuations by modulating the ideal system parameters via:
\begin{equation}
    p_{k, \text{noisy}}(t) = p_{k, \text{ideal}}(t) + \beta(t)p_{k, \text{ideal}}(t),
    \label{eq:fluctuations}
\end{equation}
where we define  $\delta p_k(t) = \beta(t)p_{k, \text{ideal}}(t)$ as the time-dependent physical fluctuation. The amplitudes $\{a_k\}$ of the underlying stochastic process $\beta(t)$ are scaled so that the root mean square (RMS) of the physical fluctuation matches the target values specified in Table~\ref{tab:noise_conditions_full}.

\begin{table}[t!]
\centering
\caption{\textbf{Noise environment parameters.} The table lists the target strength for each simulated noise source, quantified as the root mean square (RMS) of the physical fluctuation, $\delta p_k(t) = \beta(t)p_{k, \text{ideal}}(t)$. Two scenarios are presented: a \textit{Realistic Environment} with strengths derived from experimental literature, and a \textit{Strong Noise Environment} that amplifies these fluctuations by two orders of magnitude to serve as a stress test.}
\label{tab:noise_conditions_full}
\begin{tabular}{lcc}
\toprule
\textbf{Parameter Fluctuation} & \textbf{Realistic Environment} & \textbf{Strong Noise Environment} \\
\midrule
Qubit Frequency ($\delta\omega_j$) & $\sim 10^5$\,Hz & $\sim 10^7$\,Hz \\
Coupling ($\delta J/J$) & $\sim 0.001\%$ & $\sim 0.1\%$ \\
Control Amplitude ($\delta\varepsilon_{j}^{x,y}/\varepsilon_{j}^{x,y}$) & $\sim 0.01\%$ & $\sim 1\%$ \\
Control Phase ($\delta\phi_j$)& $\sim 10^{-3}$\,rad & $\sim 10^{-1}$\,rad \\
Crosstalk Variation ($\delta\alpha_{j \to i}/\alpha_{j \to i}$) & $\sim 0.01\%$ & $\sim 1\%$ \\
\bottomrule
\end{tabular}
\end{table}

We applied this stochastic model to the primary sources of noise found in fixed-frequency transmons, which can be broadly categorized into fluctuations affecting the drift and control Hamiltonians. For the drift Hamiltonian, the primary fluctuations are caused by charge noise, variations in the critical current, and microscopic two-level defects. These sources predominantly manifest as fluctuations in the qubit's frequency ($\delta\omega_j$) on the order of $10^5\,\text{Hz}$ and in the coupling strength ($\delta J/J$) of approximately $10^{-5}$ \cite{schreier_suppressing_2008, van_harlingen_decoherence_2004, schlor_correlating_2019, kristen_amplitude_2020}. Physically, these fluctuations have a $1/f$ spectral density at low frequencies, which we modeled using a $1/f$ spectrum over the range $[1\,\text{Hz}, 10\,\text{MHz}]$ and a white noise profile over $[10\,\text{MHz}, 2\,\text{GHz}]$ \cite{paladino_1_2014, tripathi_modeling_2024}. For the control Hamiltonian, we considered two main noise sources. The first is pulse imperfections, which manifest as amplitude and phase noise. Amplitude noise appears as fluctuations in the quadrature amplitudes, $\varepsilon_j^x(t)$ and $\varepsilon_j^y(t)$, with typical variations of $0.01\%$ \cite{zurich_instruments_ag_hdawg_2022, gavrielov_spectrum_2025}, while phase noise appears as a time-dependent mixing of these quadratures, corresponding to fluctuations of about $10^{-3}\,\text{radians}$ per unit amplitude \cite{gely_situ_2024}. Similar to the drift parameters, their spectral character is $1/f$ at low frequencies, transitioning to white noise at higher frequencies. This was modeled with a $1/f$ spectrum over $[1\,\text{Hz}, 100\,\text{MHz}]$ and a white noise component over $[100\,\text{MHz}, 2\,\text{GHz}]$ \cite{kumar_phase_2024, rohde_oscillator_2000, yakimov_nature_2020}. The second control noise source is classical crosstalk, where control signals for one qubit inadvertently affect another. This typically arises from how control drives are integrated into the circuit, with the magnitude of the crosstalk depending on the coupling interaction between the drive and neighboring qubits~\cite{chow_simple_2011}. To model classical crosstalk, the control term (third term) of Eqn.~\ref{Minimal_Equation_TwoQuadratureCoupledTransmons} was expanded to include terms that represent the classical coupling of one qubit’s control drive to the other qubit:
\begin{align}
    H_{\text{control}} = \sum_{\substack{i,j=1,2 \\ i \neq j}} \Bigg[ & \varepsilon_{j}^{x}(t)\left( \left(b^{\dagger}_{j}+b_{j}\right) + \alpha_{j \to i}(t)\left(b^{\dagger}_{i}+b_{i}\right) \right) \nonumber \\
    & + i\varepsilon_{j}^{y}(t)\left( \left(b_{j}^{\dagger}-b_{j}\right) + \alpha_{j \to i}(t)\left(b_{i}^{\dagger}-b_{i}\right) \right) \Bigg]. \label{Minimal_Equation_ControlCrosstalk}
\end{align}
The time-dependent crosstalk strength from the drive on qubit $j$ to qubit $i$, $\alpha_{j \to i}(t)$, was modeled as a stochastic process with a mean magnitude of $5\%$ and short-timescale variations on the order of $0.01\%$ \cite{heng_estimating_2024}, following a $1/f$ spectrum over $[1\,\text{Hz}, 100\,\text{MHz}]$ and a white noise component over $[100\,\text{MHz}, 2\,\text{GHz}]$.

We performed the robust optimization using the objective defined in Eq.~\eqref{Minimal_NoiseMinMax}, employing $N_J=3$ static samples corresponding to the nominal coupling and the error bounds ($0, \pm\Delta J_{\max}$). For the dynamic average, we utilized $L=100$ independent fluctuations $\{\boldsymbol{\beta}^{(l)}(t)\}_{l=1}^{L}$ generated via Eq.~\eqref{eq:beta_definition_appendix} at each iteration. We found that this ensemble size, combined with the inclusion of the fluctuation-free trajectory, kept the variance of the objective sufficiently low to ensure stable trust-region updates. The specific strengths of the fluctuations for the scenarios tested are detailed in Table~\ref{tab:noise_conditions_full}.

\begin{figure*}[!t]
\centering
\includegraphics[width=0.99\textwidth]{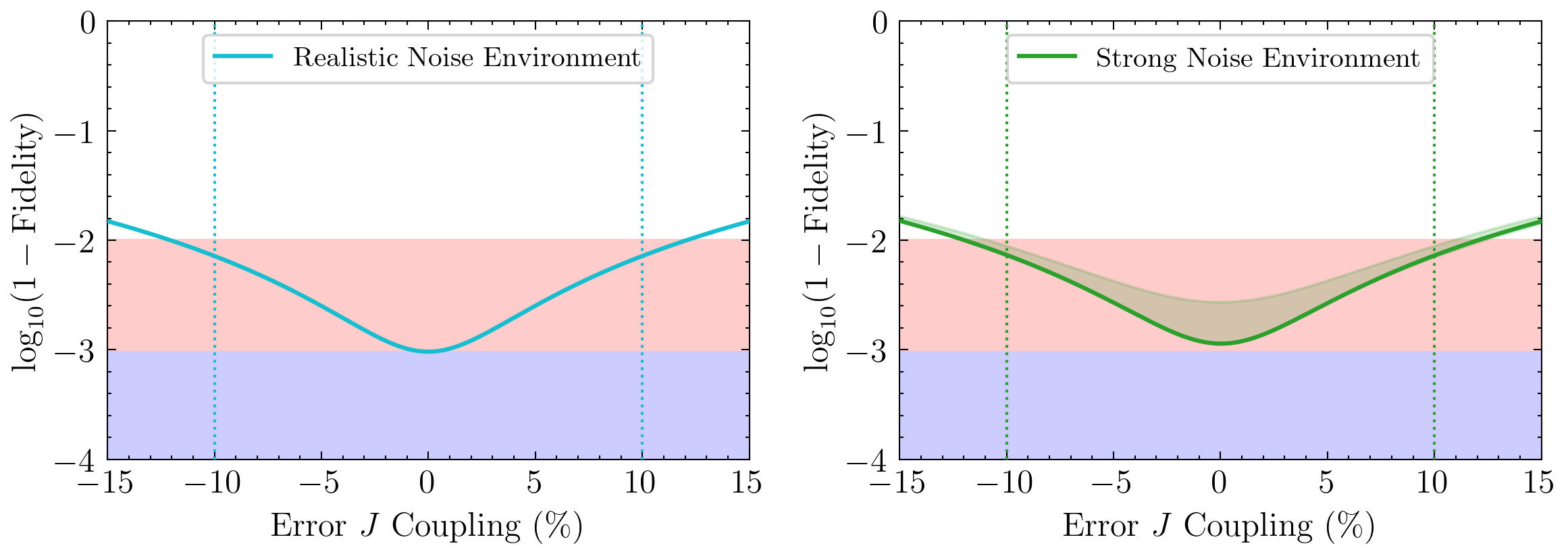}
\caption{\textbf{Performance of Cross-Resonance Gates Optimized for Simultaneous Robustness ($T=100\,\text{ns}$).} 
This figure shows the gate infidelity, $\log_{10}(1-\mathcal{F})$, of two representative $T=100\,\text{ns}$ pulses explicitly optimized to ensure simultaneous robustness against static coupling errors and time-dependent parameter fluctuations.  The left panel shows the performance of a pulse optimized for $\pm 10\%$ static error in $\Delta J/J$ alongside realistic experimental noise, achieving the desired $\mathcal{F} > 0.99$ fidelity. 
The right panel shows a pulse optimized for the same $\pm 10\%$ static error but targeting robustness against a strong-noise environment, where parameter fluctuations are two orders of magnitude larger. (Both noise environments are defined in Table~\ref{tab:noise_conditions_full}.) The solid line shows the infidelity of these robust pulses in a noise-free environment, while the shaded area represents the infidelity range across 100 unique, time dependent error realizations. The vertical dotted lines mark the $\pm$ static error range for which each pulse was specifically optimized.}
\label{Minimal_Figure_Real}
\end{figure*}

Our key finding is that we can generate sub-$100$~ns control pulses achieving fidelities $\mathcal{F} > 0.99$ that are simultaneously robust to static parameter errors (up to $10\%$) and to time-dependent errors of all parameters (up to two orders of magnitude larger than typical experimental noise). Figure~\ref{Minimal_Figure_Real} illustrates this by validating the performance of representative robust pulses against variations in the coupling $J$ across 100 independent fluctuations. This confirms that the framework can handle significant time-dependent errors without compromising static robustness.

\section{\label{Minimal_Conclusion}Conclusion}

In this work, we demonstrated that our robust optimal control framework can generate cross-resonance operations achieving fidelities of $\mathcal{F} > 0.99$ in $64$~ns with robustness against static uncertainties of up to $10\%$, and fidelities of $\mathcal{F} > 0.999$ in $71$~ns with robustness against uncertainties of up to $3\%$. This $\mathcal{F} > 0.999$ result, achieved in $71\,\text{ns}$, is nearly as fast as the non-robust time-optimal limit of $70\,\text{ns}$ \cite{kirchhoff_optimized_2018}, yet it simultaneously incorporates robustness to $\pm 3\%$ uncertainty. This combination of speed and robustness significantly surpasses previous robust strategies, which required $199\text{--}249\,\text{ns}$ to achieve similar fidelities and robustness levels \cite{allen_optimal_2017, nam_nguyen_reinforcement_2024}. This also represents a twofold improvement in gate speed compared to other experimental demonstrations of the cross-resonance gate~\cite{baum_experimental_2021, sheldon_procedure_2016, sundaresan_reducing_2020, abughanem_ibm_2025, kandala_demonstration_2021}, all while maintaining robustness against the highly uncertain qubit-qubit coupling. Furthermore, we demonstrated that this robustness extends to time-dependent errors, validating the method's effectiveness under both realistic conditions and strong-noise regimes up to two orders of magnitude beyond experimental norms.

These results are highly promising for fault-tolerant quantum computing. Fidelities of $\mathcal{F} > 0.99$ are sufficient for surface-code quantum error correction \cite{bravyi_quantum_1998, satzinger_realizing_2021, fowler_high-threshold_2009, fowler_surface_2012}, a leading error-correction strategy for superconducting devices \cite{acharya_suppressing_2023, zhao_realization_2022, andersen_repeated_2020, marques_logical-qubit_2022}. Moreover, by accepting a tighter uncertainty tolerance to reach $\mathcal{F} > 0.999$, the encoding overhead for many error-correcting codes can be dramatically reduced \cite{bravyi_high-threshold_2024, fowler_surface_2012, breuckmann_quantum_2021}.

We have further shown a fundamental trade-off in control: for any given fidelity target, there is a maximum level of uncertainty a device can tolerate, especially at short gate times. This principle is particularly relevant for noisy intermediate-scale quantum (NISQ) technologies \cite{preskill_quantum_2018, bharti_noisy_2022}, where fidelities can be relaxed to gain speed. Our results demonstrate that gate times of $51$~ns can be realized for the full range of uncertainties, provided that a lower fidelity of approximately $\mathcal{F} = 0.95$ is tolerated. This performance meets the target for many NISQ applications, and the robustness of the results ensures less variability across multi-qubit processors, which is ideal for scaling up quantum computers.

We highlight that the framework presented here are not limited to the cross-resonance gate and the methodology can be extended to any optimal control goal, such as other quantum gates. In addition, it is not limited to superconducting qubits; the methods outlined can be applied to any quantum system with any desired control objective.

While analytical explanations for pulse robustness exist, they are limited to a few idealized cases (e.g., the well-known DRAG protocol \cite{motzoi_simple_2009}) and do not account for practical constraints such as uncertainties. Numerous studies by Rabitz and colleagues have shown that the control landscape topology becomes increasingly fractured and complex under such constraints \cite{moore_exploring_2012}. This highlights the need for further dedicated research to elucidate the control landscape and pulse dynamics in these challenging scenarios. Here, we have demonstrated the use of complex-shaped pulses that achieve very high-fidelity gates, marking a significant advancement toward practical applications in quantum computing.

\section*{Funding}
E.G. acknowledges financial support from the EPSRC grant EP/T001062/1, and the EPSRC strategic equipment grant EP/L02263X/1. For R.K this material is based upon work supported by, or in part by, the U. S. Army Research Laboratory and the U. S. Army Research Office under contract/grant number W911NF2310307.

\section*{Author contributions}
E.G. conceived and supervised the project. N.G., J.L.A., and R.K. contributed to the development of the robust optimal control framework. N.G. performed the numerical simulations and data analysis. N.G. and J.L.A wrote the main manuscript text. All authors reviewed the manuscript.

\section*{Data Availability}
The datasets used and/or analysed during the current study are available from the corresponding author on reasonable request.

\newpage
\bibliographystyle{unsrt}
\bibliography{references}

\onecolumn\newpage
\appendix

\section{\label{Minimal_Pulses} Example of a Robust Cross-Resonance Control Pulse}

\begin{figure*}[htb]
    \centering
    \includegraphics[width=0.99\textwidth]{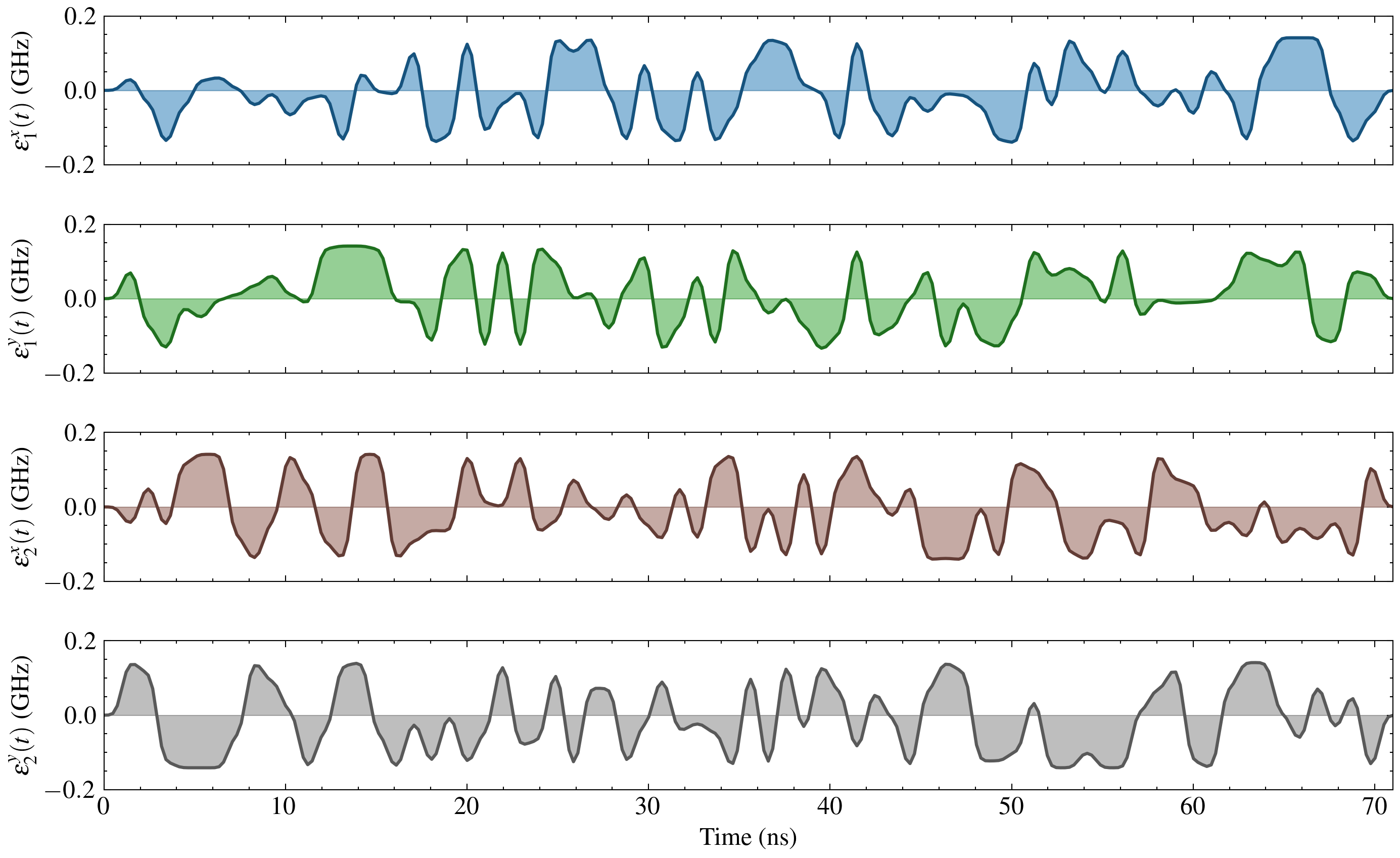}
    \caption{\textbf{Example of a Robust Cross-Resonance Control Pulse.} This pulse was obtained from the SCP optimization, achieving a fidelity of $\mathcal{F} \ge 0.999$ with a duration of $71$\,ns while maintaining robustness against $\pm 3\%$ static uncertainty in the $J$ coupling.}
    \label{Minimal_Figure_Robust_Pulse}
\end{figure*}

Figure~\ref{Minimal_Figure_Robust_Pulse} illustrates the control pulse for the shortest cross-resonance gate produced by the SCP optimization algorithm that achieves a fidelity of $\mathcal{F} \ge 0.999$ while maintaining robustness against $\pm 3\%$ static uncertainty in the $J$ coupling.

\section{\label{Minimal_Population}Population Dynamics}

\begin{figure*}[!htb]
    \centering
    \includegraphics[width=0.99\textwidth]{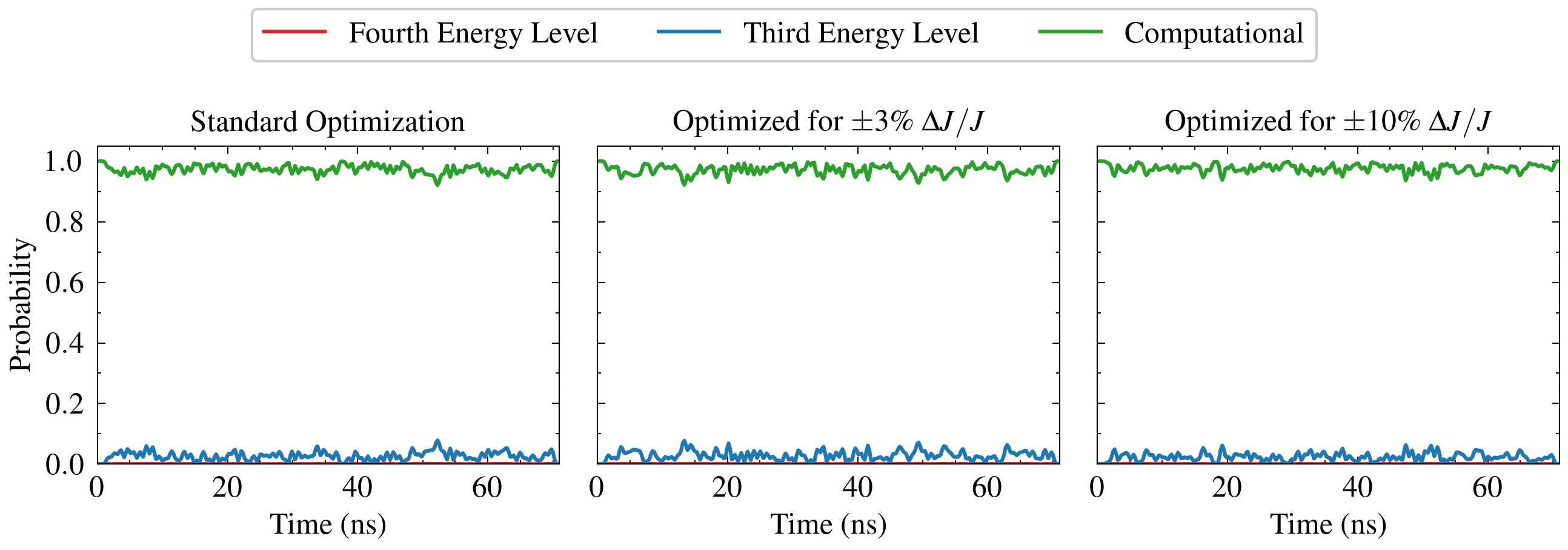}       
    \caption{\textbf{Population Dynamics of Robust Cross-Resonance Gates.} This figure shows the time evolution of populations (in the computational subspace and the third and fourth levels) during the 71\,ns cross-resonance gate obtained from SCP optimizations. The subplots illustrate the dynamics for gates optimized against varying levels of static $J$ coupling uncertainty ($\Delta J/J$): the left panel corresponds to a 0\% target (standard), the middle to $\pm 3\%$, and the right to $\pm 10\%.$}
    \label{Minimal_Figure_Leakage}
\end{figure*}

\end{document}